\documentclass[letterpaper, 10 pt, conference]{ieeeconf}  

\IEEEoverridecommandlockouts                              
\makeatletter
\let\NAT@parse\undefined
\makeatother

\usepackage[dvipsnames]{xcolor}
\newcommand*\linkcolours{ForestGreen}

\usepackage{times}
\usepackage{graphicx}
\usepackage{subfig}
\usepackage{multicol}
\usepackage{amssymb}
\usepackage{gensymb}
\usepackage{amsmath}
\usepackage{breakurl}

\usepackage{url,hyperref}
\hypersetup{
colorlinks,
linkcolor=\linkcolours,
citecolor=\linkcolours,
filecolor=\linkcolours,
urlcolor=\linkcolours}

\usepackage{algorithm}
\usepackage{algorithmic}

\usepackage[labelfont={bf},font=small]{caption}
\usepackage[none]{hyphenat}

\usepackage{mathtools, cuted}

\usepackage[noadjust, nobreak]{cite}

\usepackage{tabularx}
\usepackage{amsmath}

\usepackage{float}

\usepackage{pifont}

\newcolumntype{Y}{>{\centering\arraybackslash}X}

\usepackage[]{placeins}

\usepackage{placeins}

\usepackage{tikz}

\usepackage[framemethod=tikz]{mdframed}

\usepackage{afterpage}

\usepackage{stfloats}

\usepackage{atbegshi}
\newcommand{\handlethispage}{}
\newcommand{\discardpagesfromhere}{\let\handlethispage\AtBeginShipoutDiscard}
\newcommand{\keeppagesfromhere}{\let\handlethispage\relax}
\AtBeginShipout{\handlethispage}

\usepackage{comment}

\title{\LARGE \bf Sustaining the economy under partial lockdown: A pandemic centric approach 
}

\author{Saket Saurabh$^{1}$, Ayush Trivedi$^{1}$, Nithilaksh P. Lokesh$^{1}$, Bhagyashree Gaikwad$^{1}$
\thanks{$^{1}$All authors are researchers at the Tata Research Design and Development Centre (TRDDC), Pune, India}%
}

\begin{document}

\maketitle
\thispagestyle{empty}

\begin{abstract}
As the world fights to contain and control the spread of the Novel Coronavirus, countries are imposing severe measures from restrictions on travel and social gatherings to complete lockdowns. Lockdowns, though effective in controlling the virus spread, leaves a massive economic impact. In a country like India with 21.9\% of its population below the poverty line, lockdowns have a direct impact on the livelihood of a large part of the population. Our approach conforms to healthcare and state practices of reducing human to human contact, by optimizing the lockdown strategy. We propose resuming economic activities while keeping healthcare facilities from being overwhelmed. We model the coronavirus pandemic as SEIR dynamic model for a set of states as nodes with certain population and analyze the model output before and after complete lockdown. Social distancing that people would willingly follow, in the no lockdown situation is modeled as being influenced with the knowledge of the current number of infection by imitating Granovetter threshold model. We then provide optimal lockdown policy solutions for the duration of ten weeks using NSGA-II optimization algorithm. While there are many studies that focus on modelling the transmission of COVID-19, ours is one of the few attempts to strike a balance between number of infections and economic operations.

\end{abstract}
\section*{Disclaimer}

\noindent We wish to clearly inform that none of the authors is an epidemiologist. The current professional interest of all the authors is Operations Research, Machine learning, System Dynamics Modeling and Deep learning. Required data was present at an aggregate level. Multiple approximations were made to estimate certain values, therefore the model lacks state-level details. We are currently working on improving the model and are awaiting new details to surface online. As the testing rate varies across geographies, approximated Transmission rate has been as used by referring to WHO report.  

The toy model of interconnected states as nodes is based on multiple assumptions to analyze the effect of the pandemic and to demonstrate a possible solution to minimize infected population as well as the cost due to Covid-19. We have not attempted to model any state in a real-world context. The experiments are not intended to forecast infected population or cost. We only intend to demonstrate an optimization process that could be used with availability of state-specific data. Our model only assumes certain characteristics of each state in isolation and is not an imitation of the state.
\section{Introduction}

The first case of coronavirus is believed to be originated during the month of November in the year 2019 \cite{ma_2020} and in a matter of less than six months, the confirmed cases across the globe are over 4 Million.  In order to contain the rising morbidity and mortality, governments are pursuing policies from those that restrict travel, to policies that ensure a nationwide lockdown \cite{georgieva}. With a plummeting global economy, there is an increasing concern regarding the state of the global financial system. A recent study ascertains the economic impact of the pandemic on GDP growth, mentioning that every month of shutdown could cost approximately 2-2.5\% of global GDP growth \cite{adam_2020}. Although public health recovery is in progress, the rising number of cases of the virus and its uncontrollable transmission has caused uncertainty among private and public sectors about how or when to continue day to day operations in order to begin economic recovery. While the WHO Director-General and IMF Managing Director recently mentioned that the trade-off to save lives or save livelihoods is a false dilemma \cite{georgieva}, for getting the right balance the intervention must take place at the policy level which not only ensures restoration of public health but also incorporates strategies to mobilize economic operations by limiting enforced lockdowns within countries.

We attempt to test certain lockdown strategies by modelling the pandemic and calculating the indicators of good public health as well as indicators of the economy during partial lockdown to evaluate which lockdown strategy triumphs over complete lockdown.  A system dynamic model of the virus is used to help predict the spread of the virus in a number of states modeled as nodes with a fixed population. Certain characteristics of the population in these nodes have been modeled as parameters for the system dynamics model. These parameters are optimized to find a partial shutdown that satisfies healthcare constraints as well as economic viability.

\section{Optimal Policy}

A strategy ensuring sequential time bound lockdown to satisfy low cost to the economy as well as less infected population would constitute an optimal policy.   We first model states in which the suppression strategy is to be implemented due to rising COVID-19 cases as nodes with inter-connectivity. Each state is populated with a certain number of citizens and an initial number of infected citizens. Currently we have not included citizen level detail to model the state. Instead, we used the aggregate level data and have made assumptions in absence of data to model the pandemic as a SEIR system dynamic model.

\subsection{The Pandemic as a System Dynamics Model}
Many disease spread models are usually derived from the SEIR model which has been useful in predicting disease spread in human to human transmission when recovery fom disease takes some time. A state's population comprises of a susceptible population, infected population and a recovered population which changes with time. The population dynamics is centered around the question of how the susceptible population $S$ might get exposed to the disease and enters the bucket $E$. After an incubation period the exposed become infected, $I$ which is followed by recovery $R$ \cite{adam_2020}. Each state has health care facilities with health care capacities, $H$ which can vary from 1\% to 2\% of the state's population \cite{s_mishra_vij_sampal_2020}. We evaluate the system from a point where a few people in each state are already infected, since our interest lies more heavily on the rise in infection and it being controlled by practicing a partial lockdown. We do not intend to do a micro level observation of the disease transmission and therefore, we directly employ the transmission rate of COVID-19 from modeling as well as other related aggregate data. 

 The following notations are used to denote the variables used in the differential equation to model the virus spread.  

The notations used are in table \ref{tab:notation}.
\begin{table}[h]
    \centering
    \begin{tabular}{|l|l|}
        \hline
        Notation & Explanation\\
        \hline
         $S$ & Susceptible population \\
         $E$ & Exposed population\\
         $I$ & Infected population\\
         $D$ & Deaths\\
         $N$ & Initial population\\
         $n_{i,j}$ & Fraction of population travelling between states $i$ \& $j$\\
         $H$ & Hospital capacity\\
         $C$ & Number of serious cases\\
         $\beta$ & Transmission rate\\
         $f$ & Fatality rate\\
         $\gamma$ & Fraction of serious cases\\
         $\epsilon$ & Fraction of asymptomatic cases\\
         $\eta$ & Social distancing parameter\\
         $T_{inc}$ & Incubation period of disease\\
         $T_{prog}$ & Progression time of disease\\
         $h$ & Strain of health care system\\
         $\alpha$ & State of lockdown(0 for lockdown, 1 otherwise)\\
         \hline
    \end{tabular}
    \caption{Notations used for SEIR model}
    \label{tab:notation}
\end{table}

Equation \ref{eq:suscepible} gives the rate of change susceptible population over time. The susceptible population is directly proportional to the transmission rate $\beta$, the current susceptible population and the probability that an infected person comes into contact ($I / N$). In addition we add the rates of travel between states. The interstate travel population fraction ($n_{ij} = n_{ji}$) is assumed to be the proportional to the inverse of the distance between two states. The multi-state model we have used is very similar to the model given by \cite{arino}.

\begin{equation}
\frac{dS_i}{dt} = - \eta \beta(\alpha_i) \frac{S_i I_i}{N} + \sum_{i \neq j} n_{ji}(\alpha_i\alpha_j) S_{j} - \sum_{i \neq j} n_{ij}(\alpha_i \alpha_j) S_{i}
\label{eq:suscepible}
\end{equation}

In both the cases where people are infected or susceptible, the fraction of asymptomatic and untested people ($\epsilon$) can still find themselves likely and willing to travel to other states.  Equation \ref{eq:infected} gives the rate of infected people.

\begin{multline}
    \frac{dE_i}{dt} = \eta \beta(\alpha_i)\frac{S_i I_i}{N} + \sum_{i \neq j} n_{ji}(\alpha_i\alpha_j) E_{j} \\
    - \sum_{i \neq j} n_{ij}(\alpha_i\alpha_j) E_{i} - \frac{E_i}{T_{inc}}
    \label{eq:exposed}
\end{multline}

The Exposed component of SEIR model takes that population into account which has been exposed but not yet infectious for the duration of the incubation period ($T{inc}$) and will transition into infectious population bucket with a probability 1 at the rate of $1/T_{inc}$. The rate of change of the exposed population also contains components for interstate travel of the exposed population given in equation \ref{eq:exposed}.

\begin{equation}
\frac{dI_i}{dt} = \frac{E_i}{T_{inc}} - \frac{I_i}{T_{prog}} + \sum_{i \neq j} \epsilon n_{ji}(\alpha_i \alpha_j) I_j - \sum_{i \neq j} \epsilon n_{ij}(\alpha_i \alpha_j) I_i
\label{eq:infected}
\end{equation}

The recovery rate is directly proportional to the fraction of the population that recovers. The death rate is similarly proportional to the death rate (Equation \ref{eq:recovered}).

\begin{align}
    \frac{dR}{dt} &= (1-f_i) \frac{I_i}{T_{prog}}\\
    \frac{dD}{dt} &= f_i \frac{I_i}{T_{prog}}
    \label{eq:recovered}
\end{align}

In addition we include a component of hospital strain that influences the death rate since we expect the death rate to increase as the number of serious cases exceed hospital capacities. This component is from \cite{fiddamman_2020}.

\begin{align}
    C_i &= \gamma I_i\\
    h_i &= \frac{C_i}{H_i}\\
    f_i &= 0.04 - \frac{0.03}{1 + h_i^{0.01}}
\end{align}

The following equations help model the population dynamics in each state which acts as a  node in a network. Each state is connected with every other state and People can travel between states which are not in lockdown. Under lockdown, travel is restricted within the state and beyond the state. 

\subsection{Modeling Social Distancing}
We assume that the transmission rate is dependent both on the lockdown condition of a city as well as the social behaviour of people. This is accounted by the parameter $\eta_i$ which is dependent on the fraction of people infected in a state at any time.

\begin{equation}
    \eta_i = f(I_i)
    \label{eq:granovetter}
\end{equation}
Social distancing during no lockdown has been incorporated by taking inspiration from the Granovetter model (Equation \ref{eq:granovetter}) \cite{granovetter1978threshold}. We assume that a population takes up social distancing measures only when a certain fraction of population starts getting infected. The larger population acts collectively only when their threshold value of the fraction who are infected is exceeded. We modeled the threshold curve with the assumption that certain fraction of population (2\%) would pick up social distancing only when 0.05\% population are infected. After a certain point, an increase in infected population does not result in an increase of people following social distancing (this to model the fraction of population who might not still take the virus seriously).

\subsection{Economic cost}

A policy of suppressing the pandemic by minimising any contact can have severe economic consequences. Social distancing and lockdowns have proven to be effective, especially if applied well before large population becomes infected. The objective is to prevent over-burdening and failure of healthcare facilities which can occur when the number of infected exceeds the healthcare capacity. A study by the Imperial College in London generated estimates using their model that show the use of various control strategies to ensure a functioning health care capacity \cite{walker2020report}. Their work mentions that the impact of such suppression strategies on economic contribution could be substantial which is in agreement with the widespread understanding of the global economy.

There are various methods to model the economy. Shared variance international model uses the annual fractional change in GDP by modelling it as a normal distribution \cite{stockhammar2011probability}. However, we want to optimize policy for a certain time frame and lack state-wise cross sectional data. Therefore, we have modelled the direct costs to the economy due to the Covid-19 pandemic. The cost includes medical cost as well as the cost incurred for each revenue generating sector in the economy such as agriculture, manufactuing and service sectors. The following equations represent how the cost is calculated. 

\begin{align}
C_{med} &= c_t \times I + c_{sp} \times \gamma  I\\
C_{abs} &= p_{inc} \times I\\
C_{sd}^{sec} &= p_{inc}^{sec} \times P^{sec} \times sd^{sec}
\label{eq:costs}
\end{align}

In the absence of available sector-wise per capita incomes, we have estimated the distribution of per capita income as shown in Equation  

\begin{align}
p_{inc}^{sec} &= \frac{\text{Total income in sector}}{\text{Population working in sector}}\\
 &= \frac{\text{Total income $\times$ \% share of sector to income}}{\text{Total population $\times$ \%population working in the sector}}
\end{align}

\begin{table}[h]
    \centering
    \begin{tabular}{|l|l|}
    \hline
    Notation & Explanation\\
    \hline
         $C_{med}$ &  Medical cost\\
         $C_{abs}$ & Absenteeism cost\\
         $C_{sd}^{sec}$ & Sectoral social distancing cost\\
         $c_t$ & Cost of treatment per person\\
         $c_{sp}$ & Special life support cost\\
         $p_{inc}$ & Per capita income\\
         $p_{inc}^{sec}$ & Sectoral per capita income\\
         $P^{sec}$ & Working population in sector\\
         $sd^{sec}$ & \%population following social distancing in sector\\
    \hline
    \end{tabular}
    \caption{Notations for Equations \ref{eq:costs}}
    \label{tab:my_label}
\end{table}

The cost is calculated with the assumption that absenteeism or social distancing will lead to loss of earning. More than 50 percent of Indian population lies in emerging middle and lower income group (Data from PwC 2012 report, Table \ref{tab:econ_data}).  Due to low economic activity the earning of this group is lost as they are mostly paid on daily basis \cite{pwc}.  

\begin{table}[h]
    \centering
    \begin{tabular}{|p{2.8cm}|p{2cm}|p{0.8cm}|p{1.3cm}|}
    \hline
    \textbf{Household income/year} & \textbf{Economic class} & \textbf{\$ day per capita} & \textbf{Population (Millions) in 2010}\\
    \hline
        $<1,50,000$ & Lower & $<1.7$ & $460$\\
         $1,50,000-3,00,000$ & Emerging middle & $1.7-5$ & $470$\\
         $3,00,000-8,50,000$ & Middle & $5-10$ & $170$\\
         $>8,50,000$ & Upper middle+ & $>10$ & $460$\\
    \hline
    \end{tabular}
    \caption{Economic Data. Source: Profitable growth for the globally emerging middle, PwC, 2012 \cite{pwc}}
    \label{tab:econ_data}
\end{table}


\subsection{Optimization}
The economic cost increases during no lockdown due to a rise in infected population which leads to a rise in medical cost, whereas during lockdown the cost will also increase due to loss of work in different economic sectors. An attempt to minimize the cost by using an optimal policy has been done. The generated optimal policy will provide a solution to allow economic activities only when the chance of rise in infection is low, thus that would allow lower medical costs as well as an escape from loss of work.

To generate an optimal policy we use the Non-dominated Sorting Genetic Algorithm -2 (NSGA-2) for optimization \cite{deb2000fast}. Given that the dynamics of the SEIR model is higly non-linear, the NSGA-2 algorithm is ideal for optimization. Moreover, this algorithm provides the pareto front when there are more than one objective functions. The objectives used here are:
\begin{enumerate}
    \item Cost to the economy
    \item Average number of infected people over the period of policy
\end{enumerate}
For each candidate policy which represents one individual in the genetic algorithm (GA), we use the system dynamics model to predict the number of infected people and hence the cost to the economy. These objectives act as the fitness values needed for the GA.
The lockdown policy is kept constant for a period of time (say 1 week) and is determined statewise. We assume lockdown is represented by 0 and no-lockdown as 1. Thus, the optimization is done over the domain of binary matrices of size $S \times T$ where, $S$ is the set of states and $T$ is the time periods with each element determining the state of lockdown at a given state and period of time.

\section{Experiments}

In this study, we obtain an optimal policy for 9 states for a period of 10 weeks with intial infection period of 22 days where no lockdown is imposed.  Optimization is done over the average number of infected population and the total economic cost over 300 days. We collected data for nine states in India and selected these states due to the variation of workforce distribution in different sectors. These states have been modeled with the assumption that it is free from the effect of other economic factors such as international trade, production in other factors etc. We do not intend to model an actual state, our interest is limited to study of COVID-19 in the presence of few factors of interest in order to demonstrate the generation of an optimal policy. Some details of the states are in Table \ref{tab:state_data}

\begin{figure*}[tbh!]
\begin{tabular*}{\textwidth}{|m{2.7cm}|c|c|c|ccc|ccc|}
\hline
    \multicolumn{1}{|c|}{State} & \multicolumn{1}{c|}{Population} & \multicolumn{1}{c|}{Initial Infected} & \multicolumn{1}{c|}{Income(INR)} &	\multicolumn{3}{c|}{Economic Sector} & \multicolumn{3}{c|}{\% working in lockdown}\\
    &&&&Agri.&Manuf.&Serv.&Agri.&Manuf.&Serv.\\
    \hline
    Andhra Pradesh & 49577103 &	10 & 52814 & 19.04 & 57.34 &	23.61 &	100 & 77 & 66\\
    Delhi &	16787941 &	10 & 129746	 & 0.64	& 90.16	& 9.2 &	100	& 77 &	66\\
    Jharkhand &	32988134 &	10 & 27132 & 16.65 &	16.8 &	45.86 &	100	& 77 &	66\\
    Karnataka &	61095297 &	10 &	52191 &	13.5 &	59.42 &	27.08 &	100 & 77 &	66\\
    Maharashtra &	112374333 &	10	& 74027 &	7.54 &	64.03 &	28.43 &	100	& 77	& 66\\
    Punjab	& 27743338	& 10	& 62605 &	20.84 &	50.86 &	28.3 &	100	& 77 &	66 \\
    Tamil Nadu &	72147030 &	10 &	63547 &	7.28 &	63.7&	29.02 &	100	&77 &	66\\
    Uttar Pradesh &	199812342	& 10	& 23392 &	22.19 &	56.63&	21.19&	100&	77&	66\\
    West Bengal	& 91276115&	10&	41837&	16.6&	65.05&	18.35&	100&	77&	66\\
    \hline
\end{tabular*}
\vspace{\baselineskip}
\captionof{table}{Economic data of 9 states used in experiments. The income is the average per capita income per year for the state across three sectors: Agriculture, Manufacturing and Services. We calculate the cost for each state using sector wise earning and per capita earning of individuals in each state. The data tables were available at the Planning Commission: Government of India website \cite{planning}.}
\label{tab:state_data}
\end{figure*}

The treated fraction rate is set uniformly at 20\% of the infected population collected from currently available data \cite{s_mishra_vij_sampal_2020} and the untreated fraction rate is 10\% of the infected population. The studies and observations around COVID-19 specify in general that the average disease incubation period is around 5 days while the average disease infection period (progression time) is about 14 days; the data has been taken from the WHO\cite{coronavirus}. 

In each state the fraction of asymptomatic (and untested) cases, $A$ is assumed to be 20\% of the infected population of the state and the fraction of serious cases is also taken as 20\% percent of the infected population. We assume that the asymptomatic and untested population are able to move freely between cities during a no lockdown period. Using the current data available, the transmission rate is on average 2.2 during no lockdown and is assumed to be 1 during the time when lockdown is in effect.

\subsection{Effect on policy on infections}
\begin{figure}[h!]
     \centering
     \includegraphics[width=0.8\linewidth]{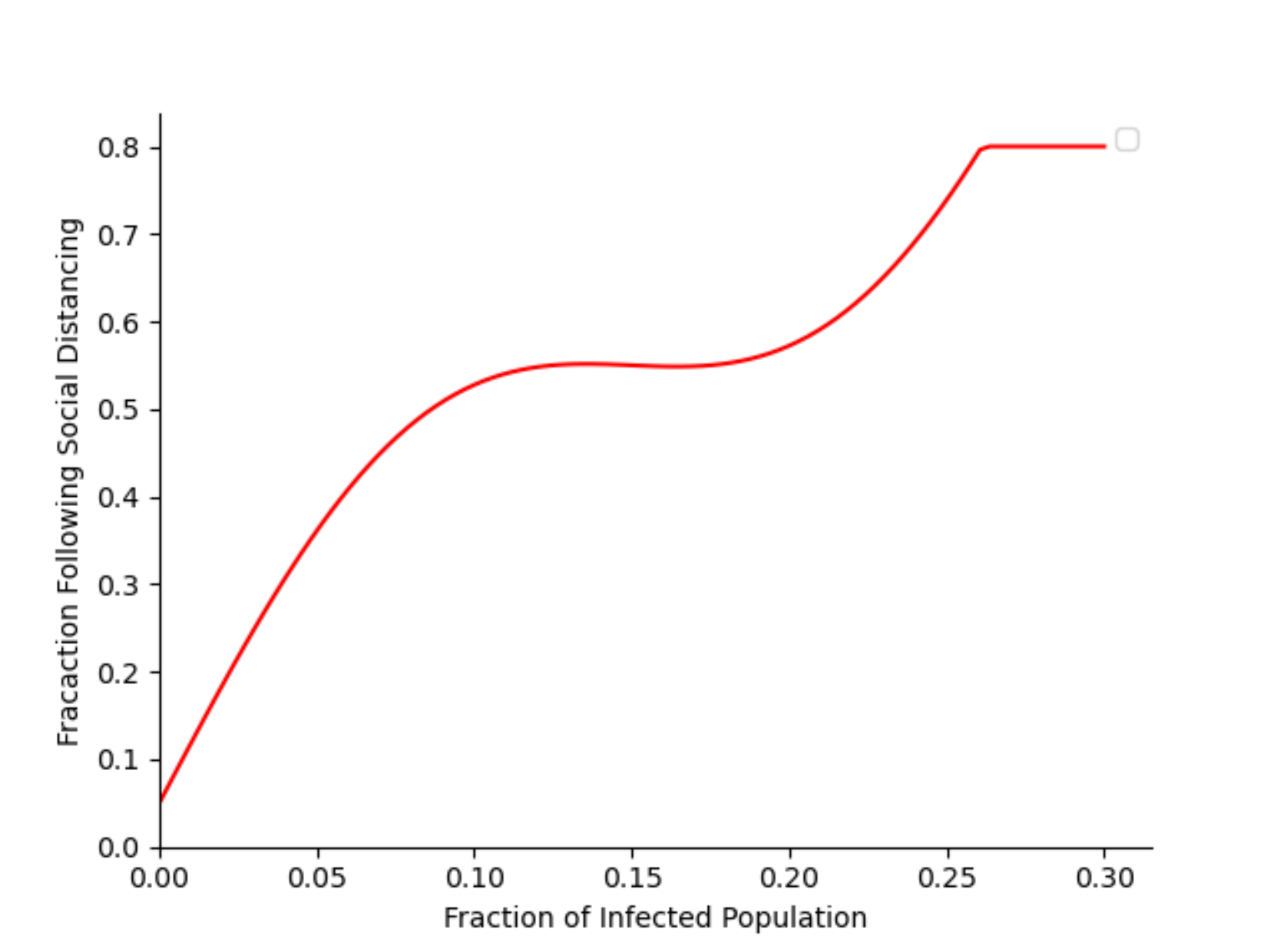} 
     \caption{Threshold curve -1: This represents a gradual adoption of social distancing as infections rise with a more steep increase as infections reach 20\% of the population. The social distancing fraction is capped at 0.8.}
     \label{fig 1a}
 \end{figure}

To study the effect of social distancing, we use two models of
the social distancing factor $\eta$. Threshold curve - I in figure \ref{fig 1a} represents a gradual adoption of social distancing as the number of infections begin to rise.
We assume that a relatively high threshold of 20\% of infected population, fraction of people adopting social distancing rises quickly, as people realize the seriousness of the pandemic.

\begin{figure}[h!]
     \centering
     \includegraphics[width=0.8\linewidth]{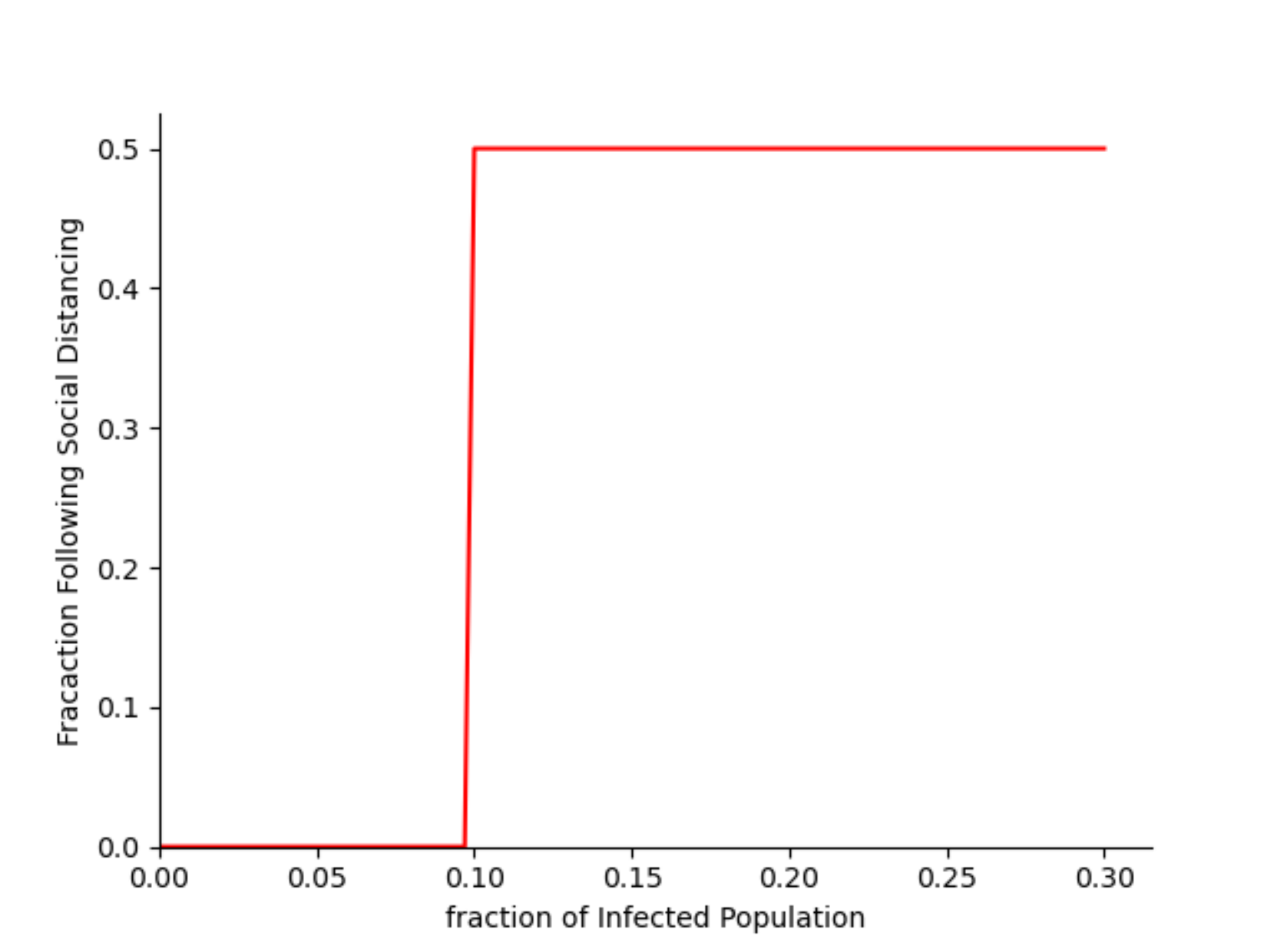} 
     \caption{Threshold curve – 2: A sudden rise in adoption of social distancing is modelled by using a step function.}
     \label{fig 1b}
 \end{figure}

In figure \ref{fig 1b}, people suddenly follow social distancing after a certain threshold and is modeled as an step function. The social distancing factor is taken as,
$\eta = 1 - \text{fraction of population following social distancing}$.

{}
 
 {}

\begin{figure*}
\begin{multicols}{2}
    \includegraphics[width=\linewidth]{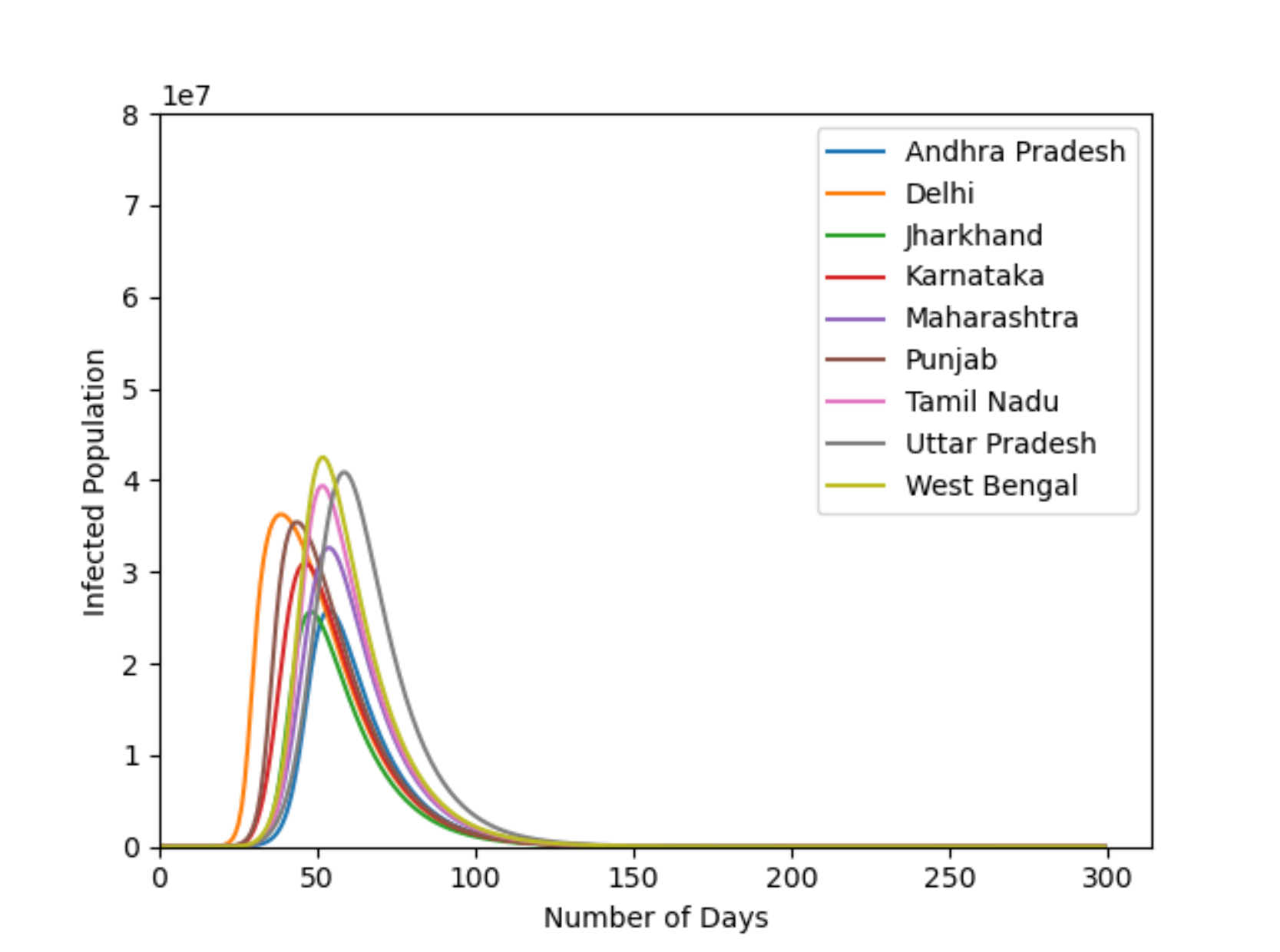}\par
    \caption{No lockdown, no Social Distancing - infections peak early and overwhelm the healthcare capacity.}
    \label{fig 2a}
    \includegraphics[width=\linewidth]{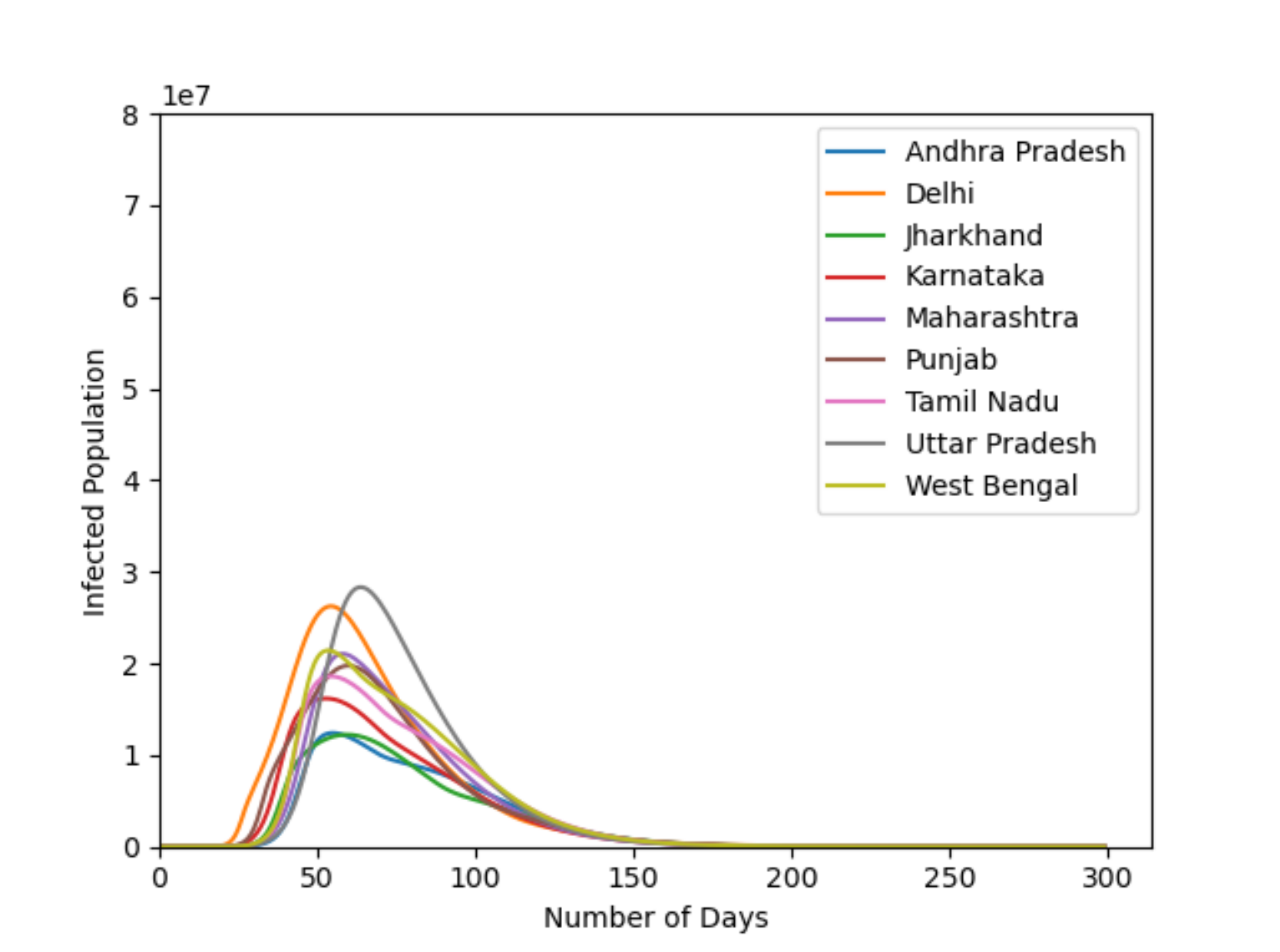}\par
    \caption{No lockdown but with social Distancing in
effect according to Threshold Curve - I}
    \label{fig 2b}
\end{multicols}
\begin{multicols}{2}
    \includegraphics[width=\linewidth]{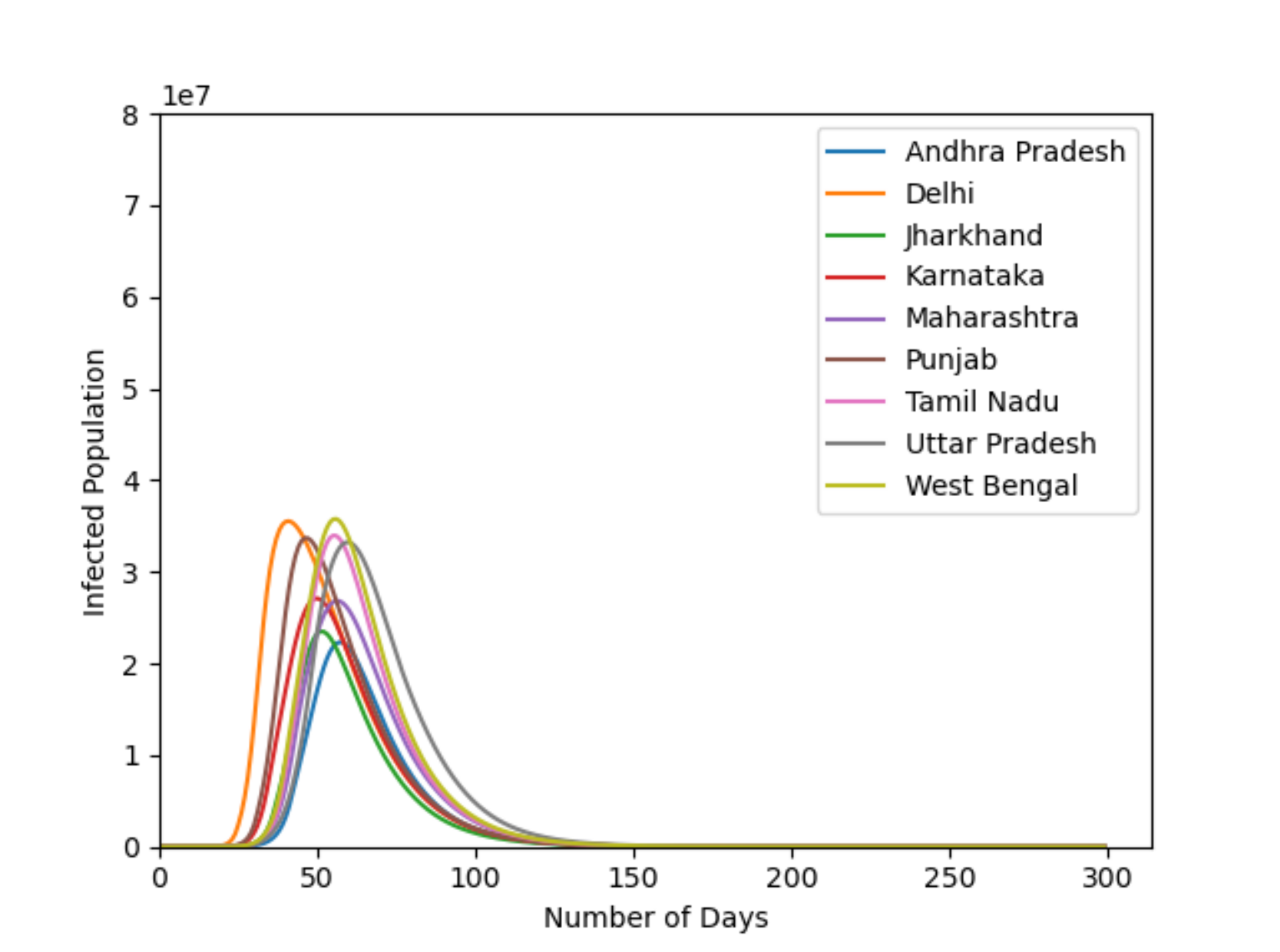}\par
    \caption{No lockdown but with step social Distancing in effect according to Threshold Curve - II}
    \label{fig 2c}
    \includegraphics[width=\linewidth]{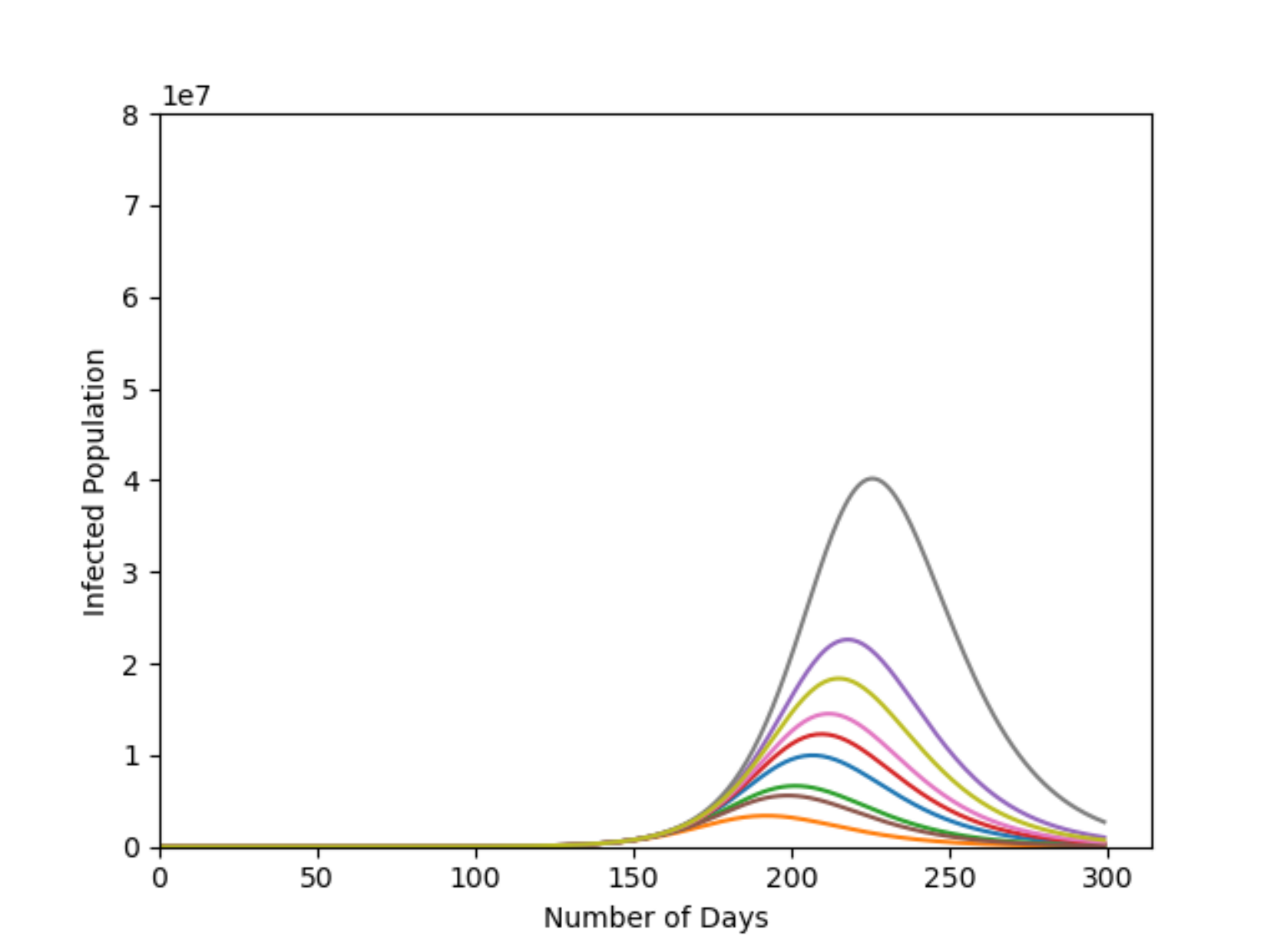}\par
    \caption{Complete Lockdown}
    \label{fig 2d}
\end{multicols}

    \label{fig 2}
\end{figure*}

We generate results for a period of 300 days for a certain policy  decision to demonstrate the effect of lockdown and social distancing on the infection curve. Without a complete lockdown, the infection peaks early (Figure \ref{fig 2a}) but applying social distancing causes the curve to flatten (Figures \ref{fig 2b} and \ref{fig 2c}). On the other hand, when a complete lockdown is in effect the curve flattens completely.  The Figure \ref{fig 2d} shows a situation where people might not conform to social distancing measure during a no lockdown policy. They only follow social distancing (Step social distancing) when a collective threshold of infected population is reached. Even though people follow social distancing after knowing that a large population is infected, the delay in intervention leads to early rise in overall infected population. 

\begin{figure*}
\begin{multicols}{2}
    \includegraphics[width=\linewidth]{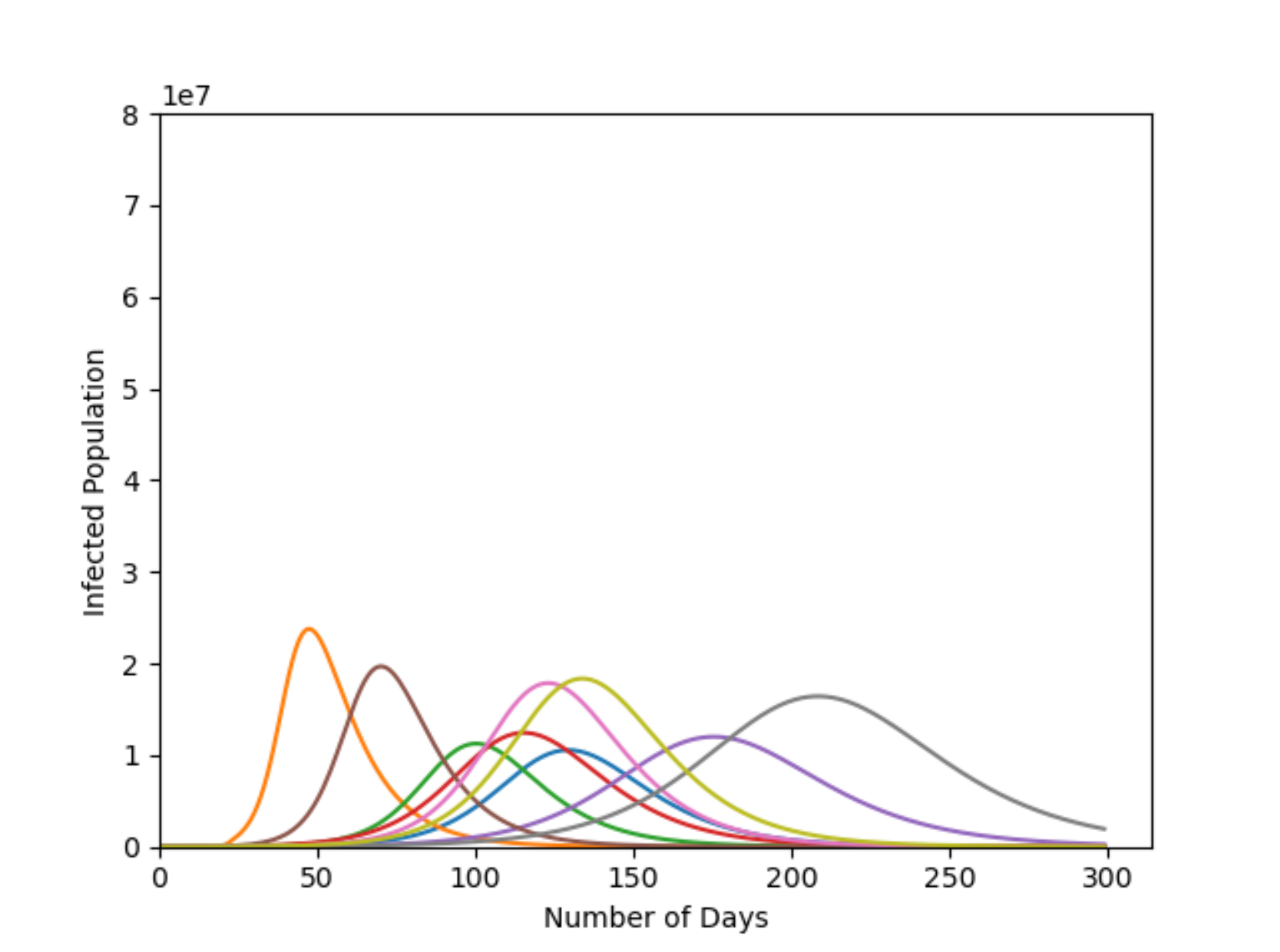}\par
    \caption{No lockdown for 22 days, lockdown for 150 days}
    \label{fig 2e}
    \includegraphics[width=\linewidth]{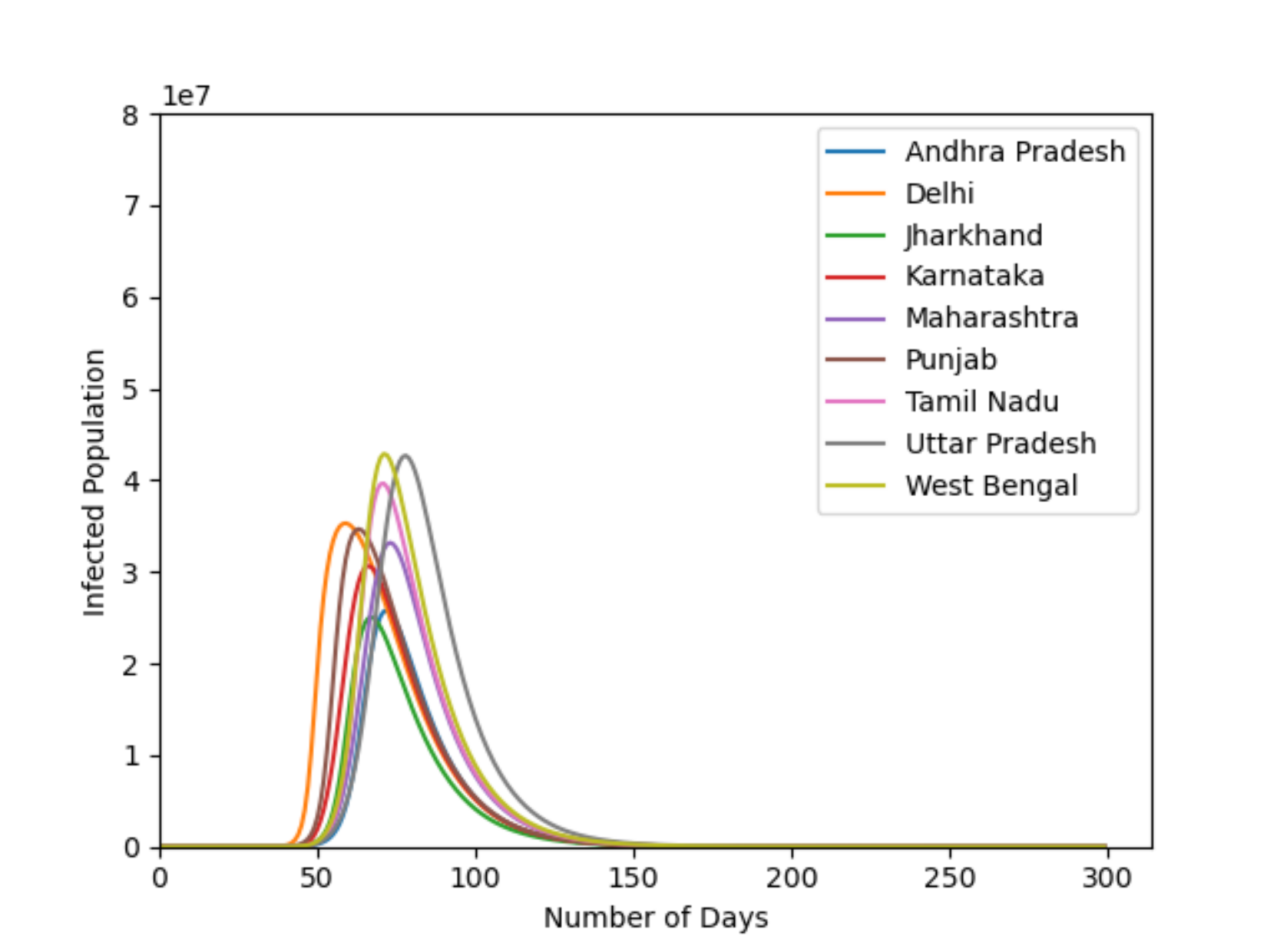}\par
    \caption{No lockdown for 22 days, then lockdown for 22 days}
    \label{fig 2f}
\end{multicols}
\caption*{\textbf{Effect of policy}}
\end{figure*}

Figures \ref{fig 2e} and \ref{fig 2f} show the effect of change in lockdown policy. The figure \ref{fig 2e} shows the number of infected people, if the lockdown is implemented after 22 days from the first infection and then removed after 150 days. This shows a flatten curve but also shows a rise in infected population later in time. The curve \ref{fig 2f} shows infected population if the lockdown is implemented for 22 days and then removed after 22 days: this leads to a sharp early rise in the number of infected population.

\begin{figure*}
\begin{multicols}{2}
    \includegraphics[width=0.9\linewidth]{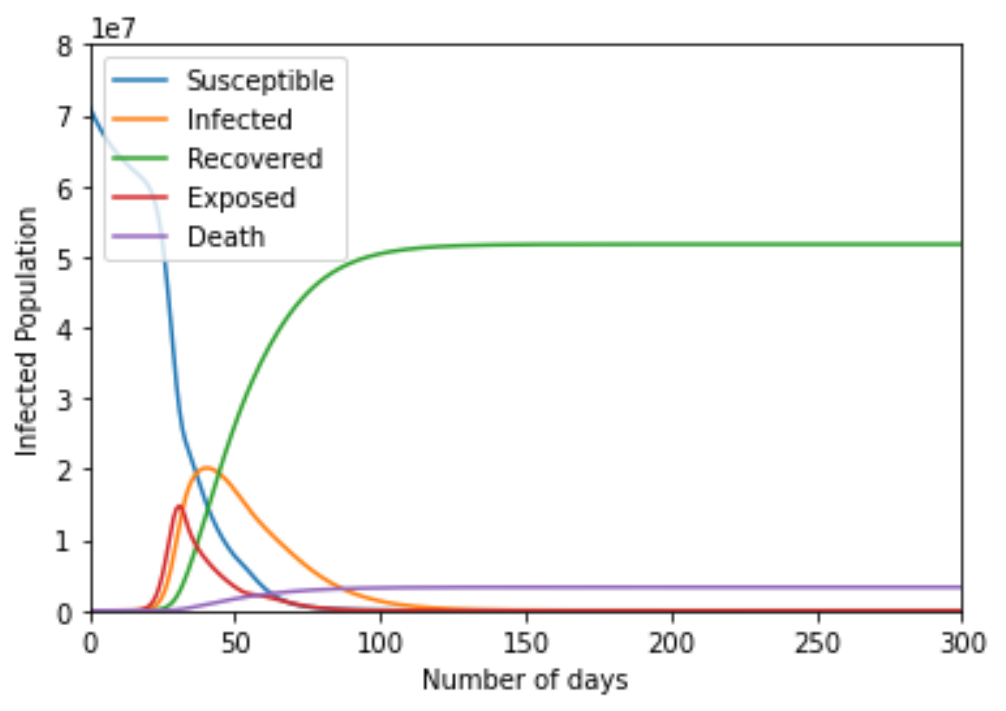}\par
    \caption{No lockdown}
    \label{fig 3a}
    \includegraphics[width=0.9\linewidth]{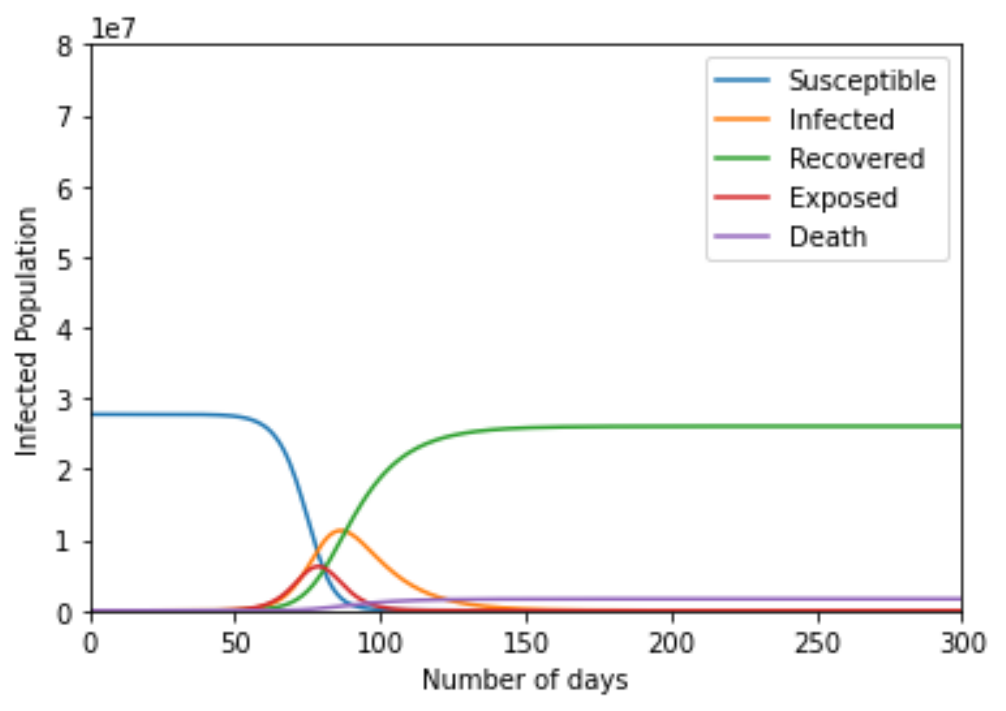}\par
    \caption{Lockdown}
    \label{fig 3b}
\end{multicols}
\caption*{\textbf{Effect of lockdown for Andhra Pradesh}}
\label{fig 3}
\end{figure*}

The figures \ref{fig 3a} and \ref{fig 3b} show effect of lockdown for the state of Andhra Pradesh with the curves for susceptible, infected, recovered populations and deaths.

\subsection{Effect of policy on Economy}

\begin{figure*}
\begin{multicols}{2}
    \includegraphics[width=\linewidth]{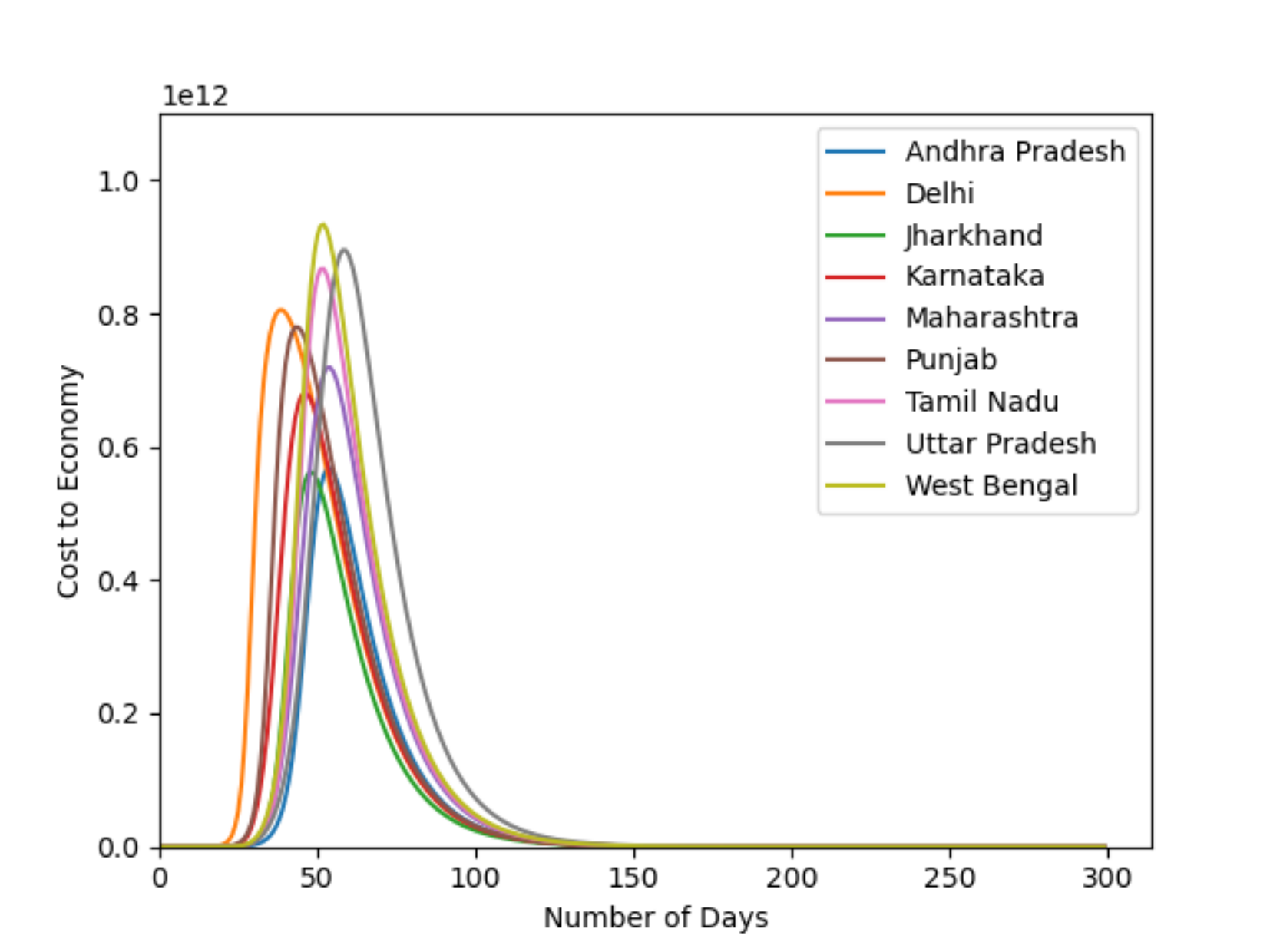}\par
    \caption{No lockdown, no Social Distancing}
    \label{fig 4c}
    \includegraphics[width=\linewidth]{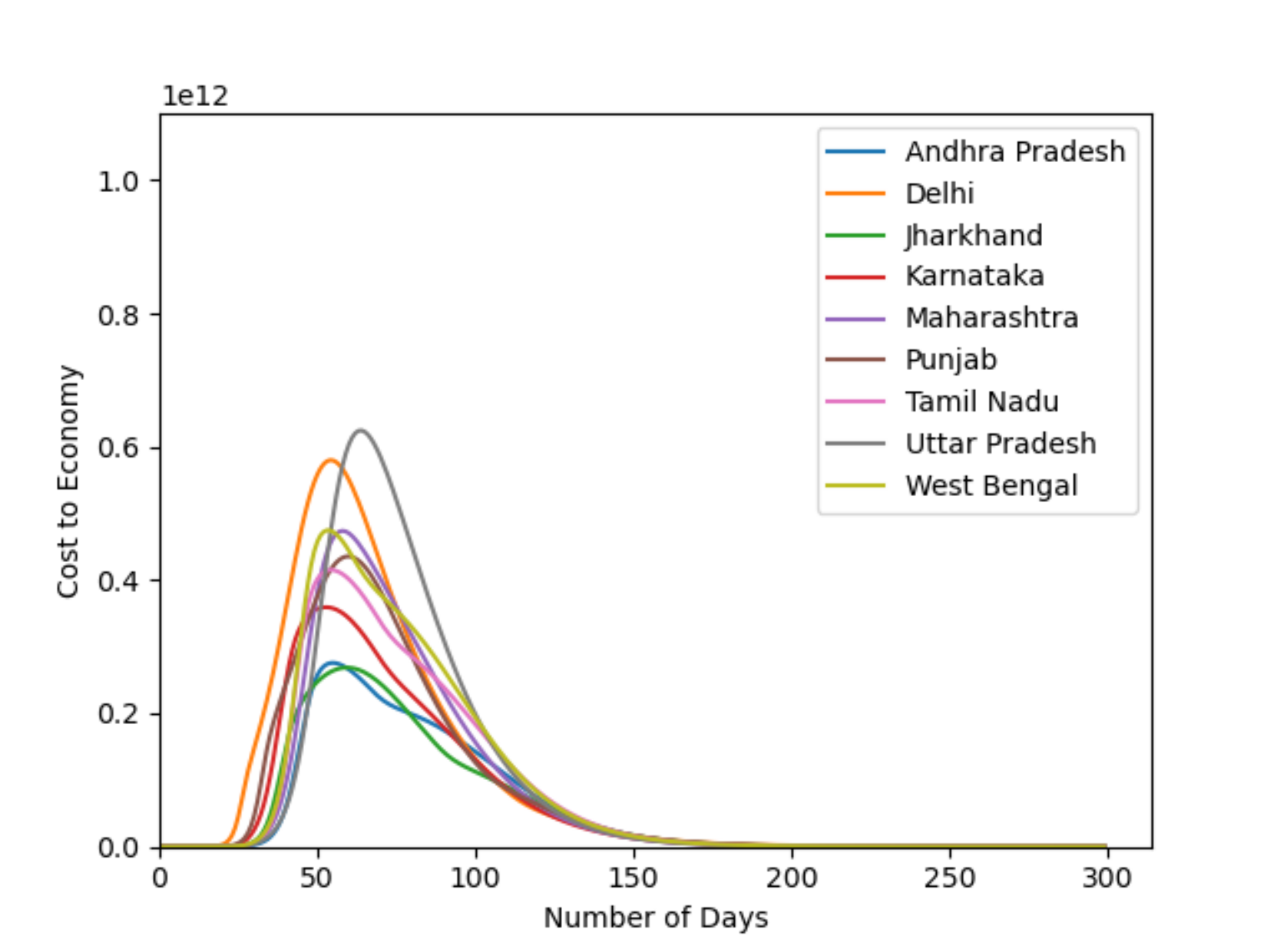}\par
    \caption{No lockdown but with social Distancing in
effect according to Threshold Curve - I}
    \label{fig 4a}
\end{multicols}
\begin{multicols}{2}
    \includegraphics[width=\linewidth]{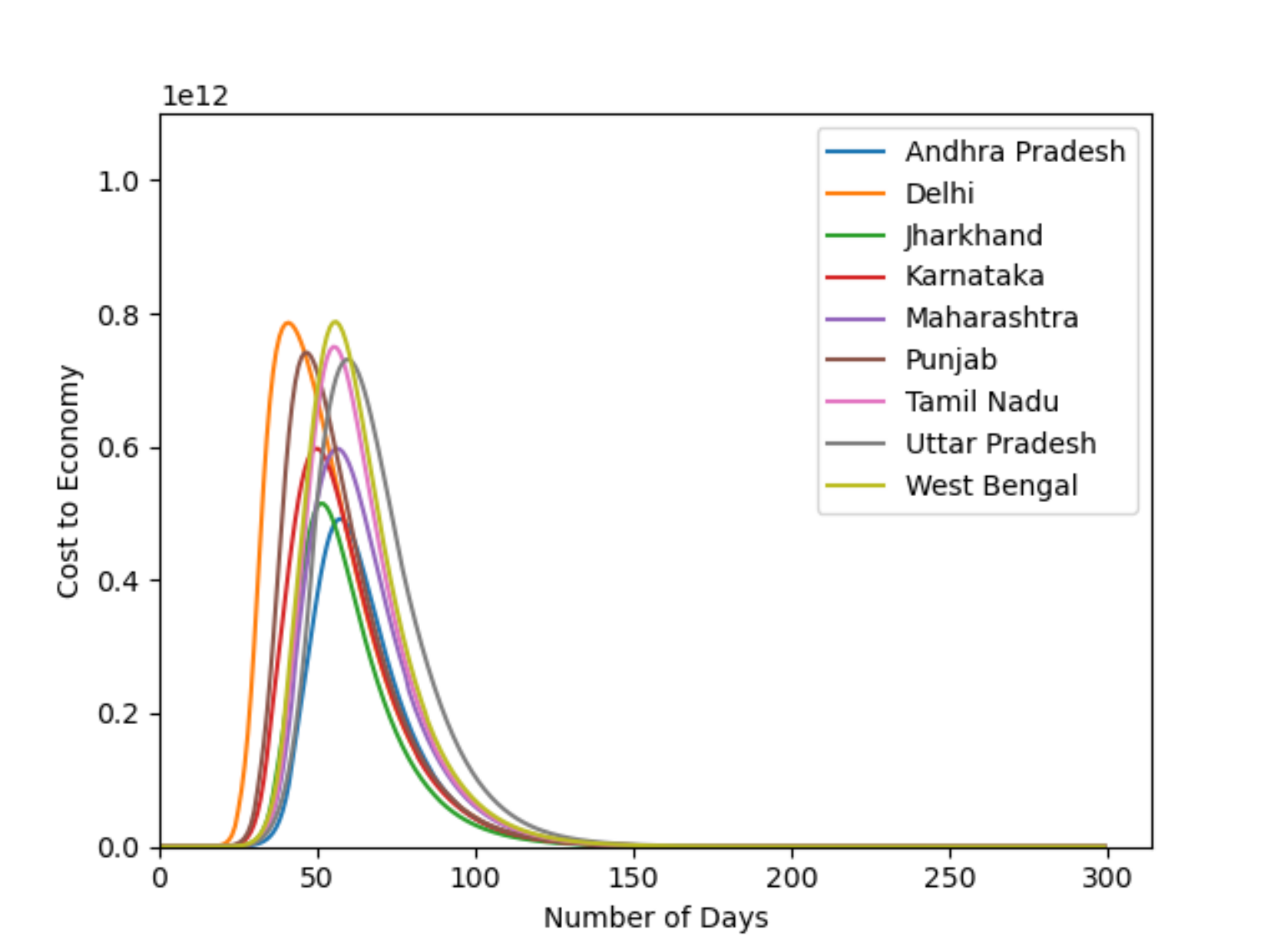}\par
    \caption{No lockdown but with step social Distancing in effect according to Threshold Curve - II}
    \label{fig 4d}
    \includegraphics[width=\linewidth]{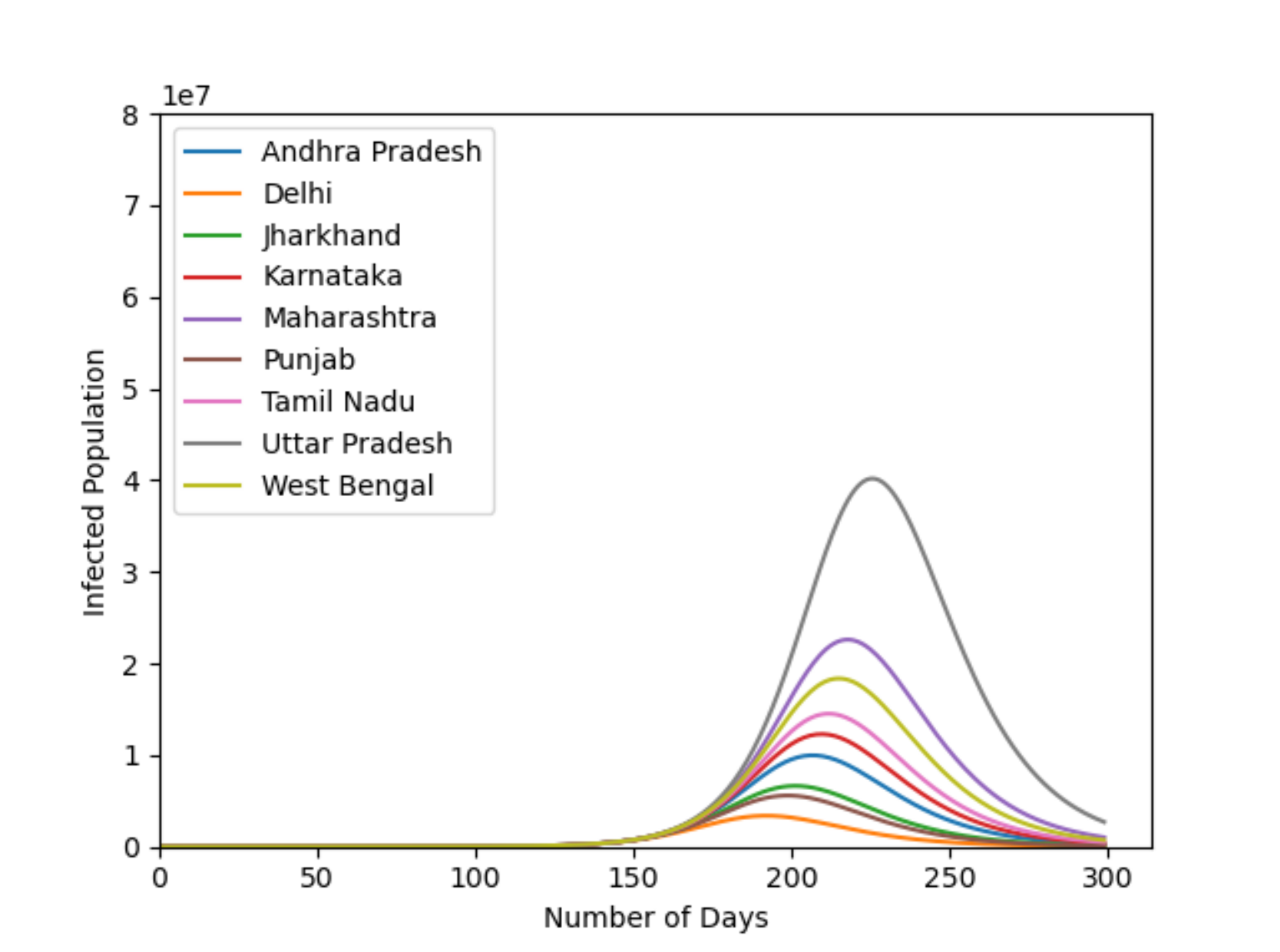}\par
    \caption{Complete Lockdown}
    \label{fig 4b}
\end{multicols}
\caption*{\textbf{Effect of policy on the cost to economy}}
    \label{fig 4}
\end{figure*}

The effect of policy on the cost to economy is simulated for three hundred days in figures \ref{fig 4a} - \ref{fig 4d}. The most anticipated effect of lockdown occurs at the time of infection peak, where more people get infected leading to rise in cost. When no suppression there is a positive correlation of COVID-19 cases and the GDP, which is found by the research group (Zhang et.al, 2020) \cite{zhang2020epidemic}.  

Figures \ref{fig 4a}  and \ref{fig 4d} shows the cost during no lockdown with social distancing in effect while figure \ref{fig 4b} shows the cost during lockdown whereas the figure \ref{fig 4c} shows the cost with no suppression strategy such social distancing in effect. Clearly, not imposing a lockdown 
imposes large costs in a quick span of time primarily due to the large medical costs incurred. On the other hand a full lockdown also imposes large economic cost due to the loss of income for people who cannot work due reduced work capacity of different sectors of the economy. 

Hence, an optimal policy must be sought in order to reduce both medical and absenteeism costs while keeping the number of infections low.

\subsection{Optimal policies}
For our policy optimization experiment, we take the nine states mentioned in table \ref{tab:state_data} with transmission rates as $\beta = 1$ for no lockdown and $\beta=0.1$ during lockdown. \textcolor{blue}. 

The NSGA-2 algorithm is used to derive the pareto front against competing objectives of minimizing total cost to the economy and the average number of infections. The optimal policy is generated over 100 generations with a population of 50 for the Genetic Algorithm. 

We initialize the simulation for 30 days with seed infections in each state as 10 infected individuals. The optimal policy is generated in week wise blocks for 10 weeks. Each policy decision in each state for lockdown is thus maintained for 7 successive days. The rest of the parameters are the same as in the previous sections. The policy generated is a week schedule containing lockdown status of each state for that week. 

\begin{figure*}
    \centering
    \includegraphics[width=0.75\linewidth]{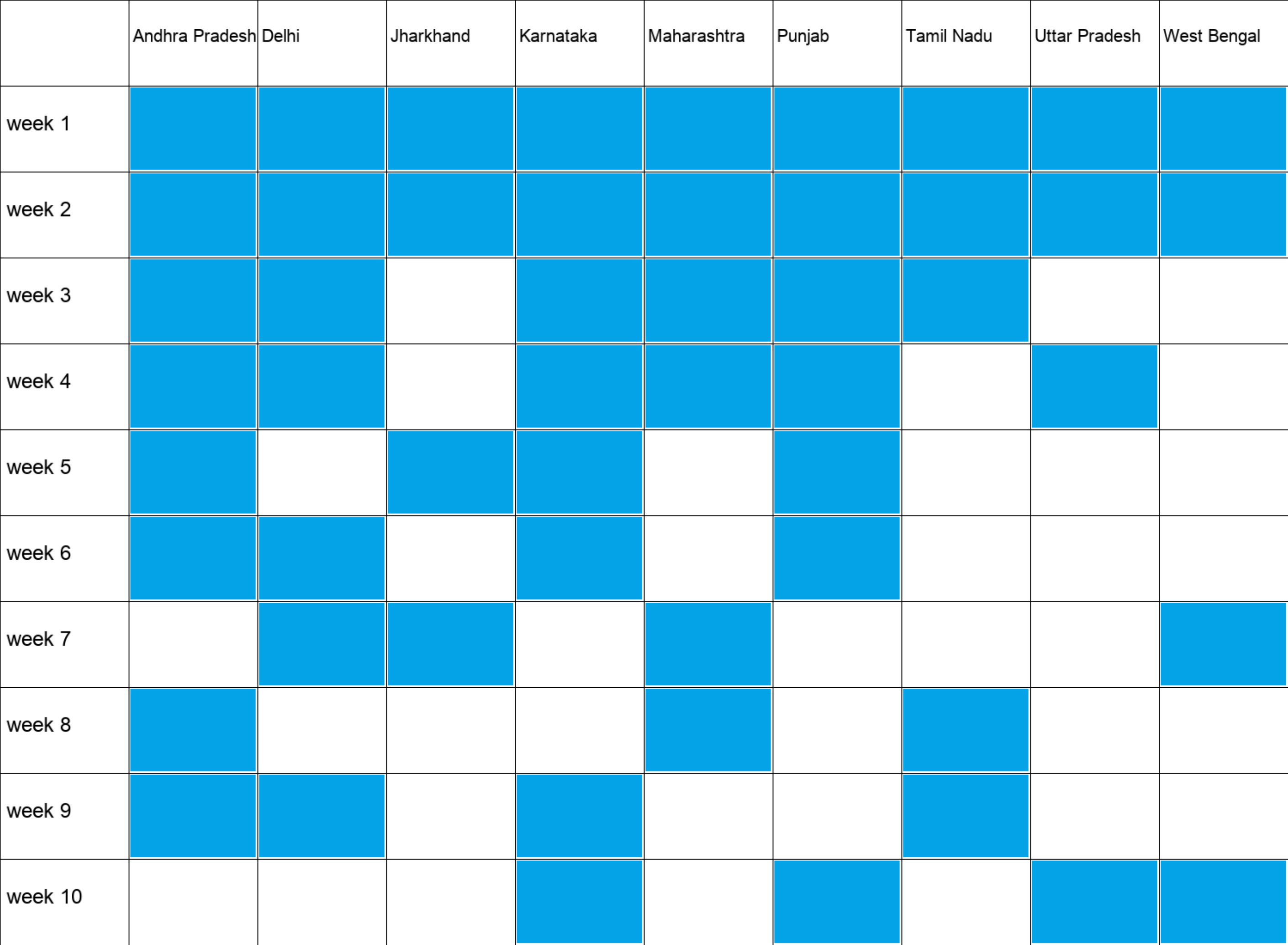}\par
    \caption{Optimal policy for Threshold curve - I}
    \vspace{1cm}
    \label{fig 5a}
    \includegraphics[width=0.75\linewidth]{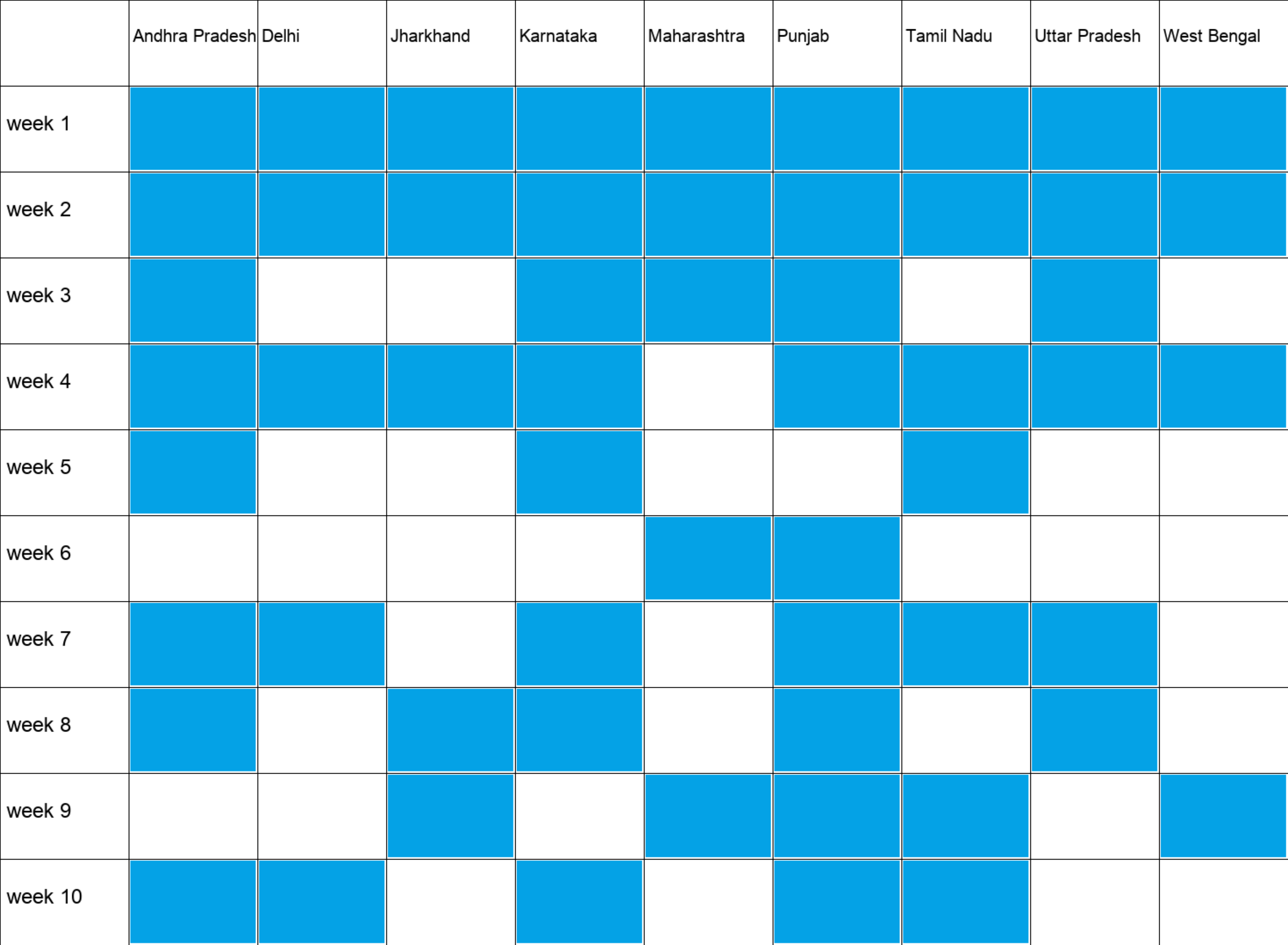}\par
    \caption{Optimal policy for Threshold curve - II}
    \label{fig 5b}
\caption*{\textbf{Optimal lockdown schedules for 10 weeks for 9 states. Filled rectangles represent lockdown while empty rectangles represent no lockdown.}}
    \label{fig 5}
\end{figure*}

An optimal lockdown strategy involves removing a suppression strategy to engage safe economic activity in a state with an extremely low infected population. Each solution of the pareto front (Figure \ref{fig 5}) shows the lockdown status for all states at different locations for a time period 10 weeks. Lockdown status remains constant for the period of first 2 weeks which is complete lockdown followed by altering the lockdown status of few states. The solutions vary from minimizing people getting infected to minimizing ecnonomic cost. The choice of selection could be made to conform to other factors in place such as hospital capacity, fraction of population that might not need treatment, etc. We demonstrate only one of the pareto-optimal policies since it was observed that all the candidate pareto-optimal solutions have almost indistinguishable objective values.

\begin{figure*}[t]
\begin{multicols}{2}
    \includegraphics[width=\linewidth]{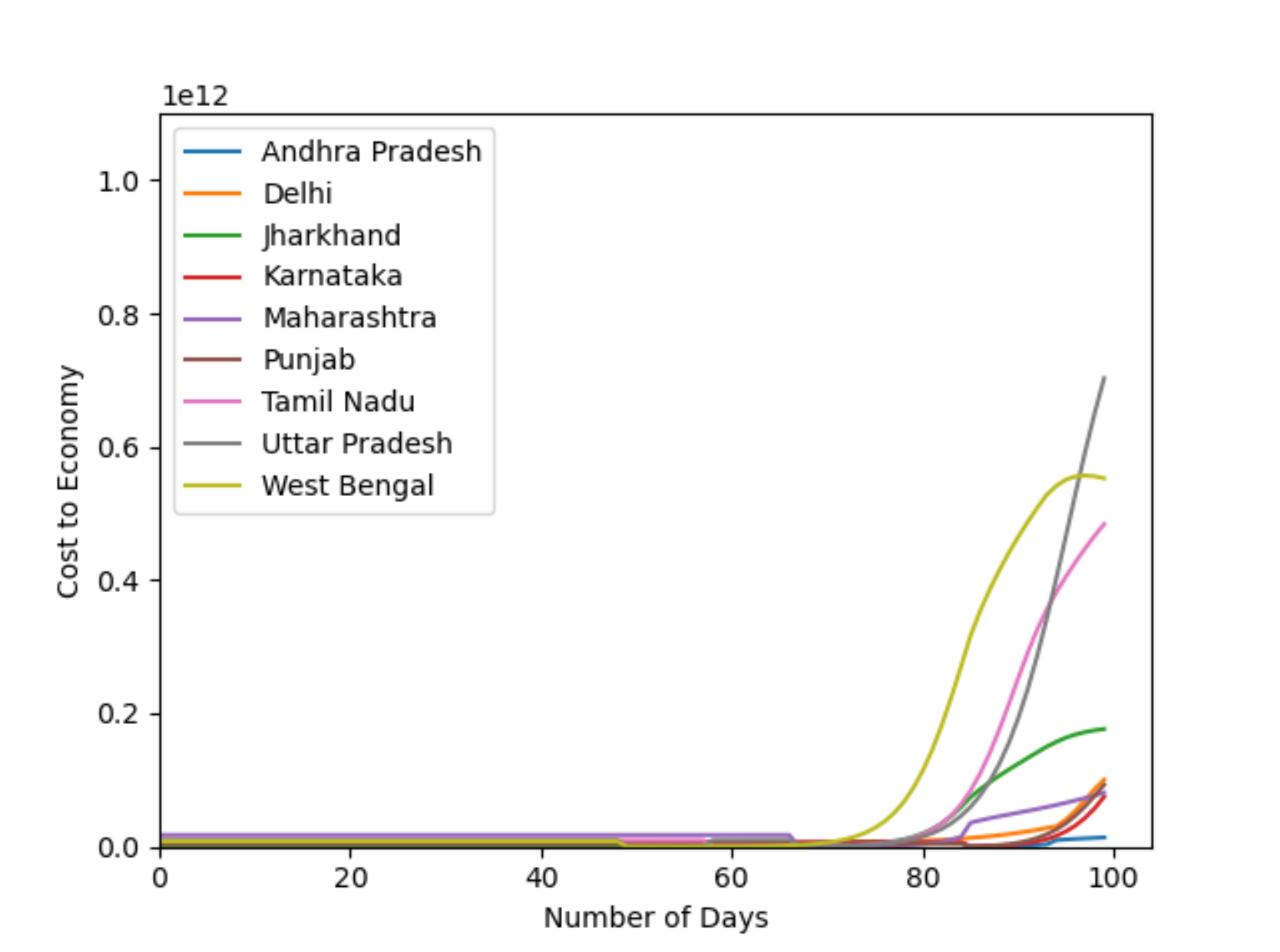}\par
    \caption{Threshold curve - I}
    \label{fig 6a}
    \includegraphics[width=\linewidth]{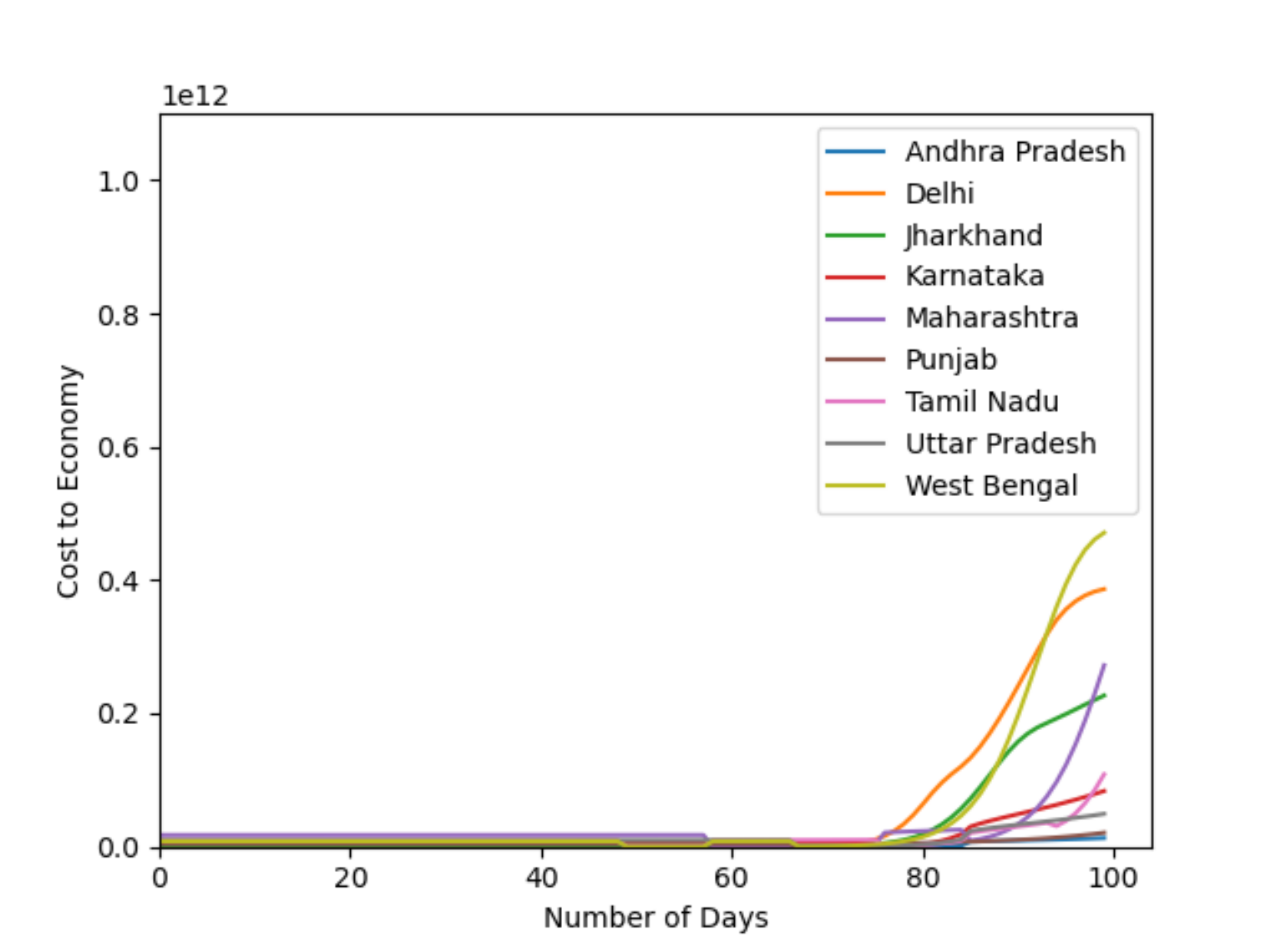}\par
    \caption{Threshold curve - II}
    \label{fig 6b}
\end{multicols}
\caption*{\textbf{Cost to economy of different states for 100 days using optimal policy}}
    \label{fig 6}
\end{figure*}

\begin{figure*}[t]
\begin{multicols}{2}
    \includegraphics[width=\linewidth]{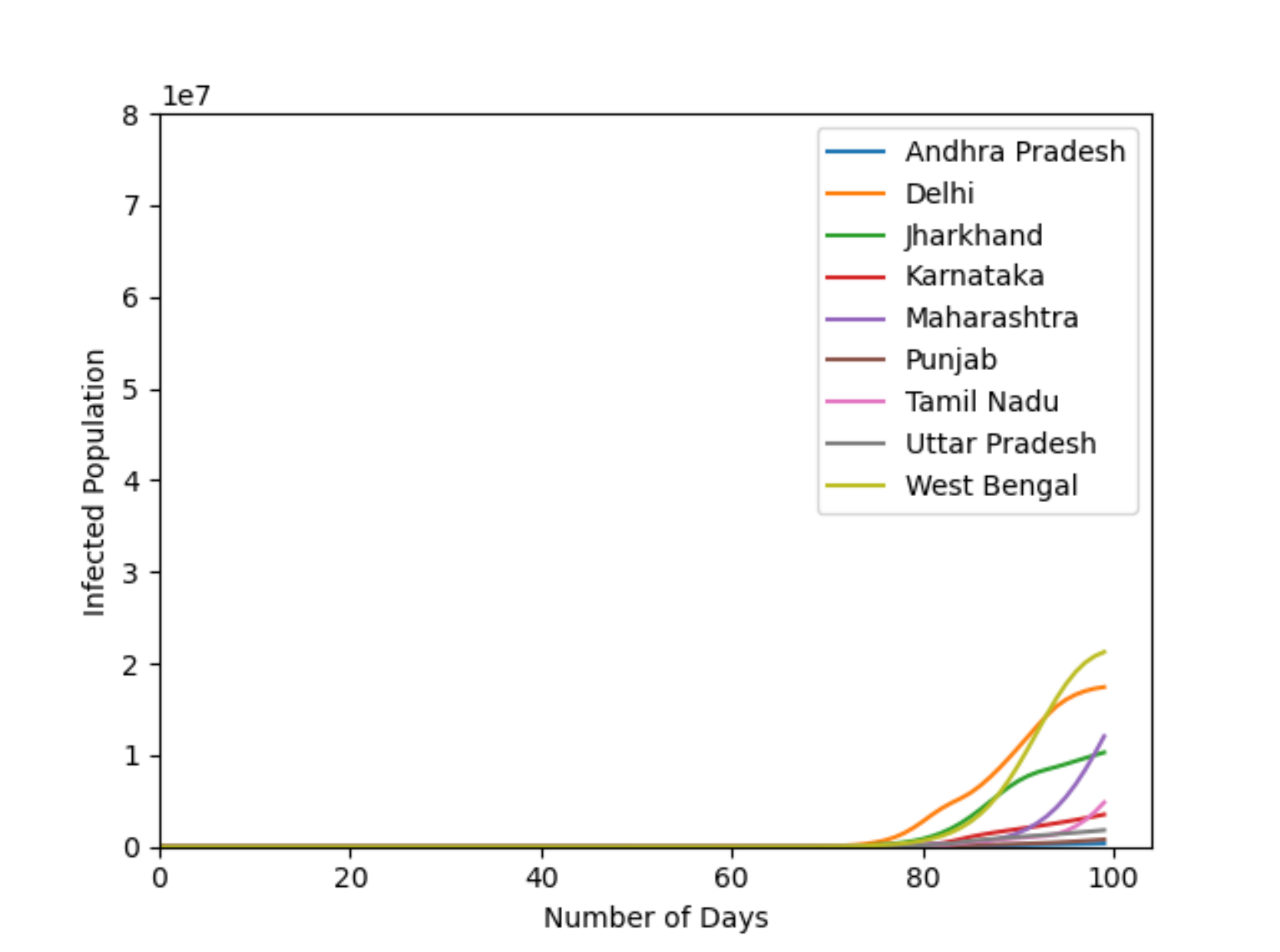}\par
    \caption{Threshold curve - I}
    \label{fig 7a}
    \includegraphics[width=\linewidth]{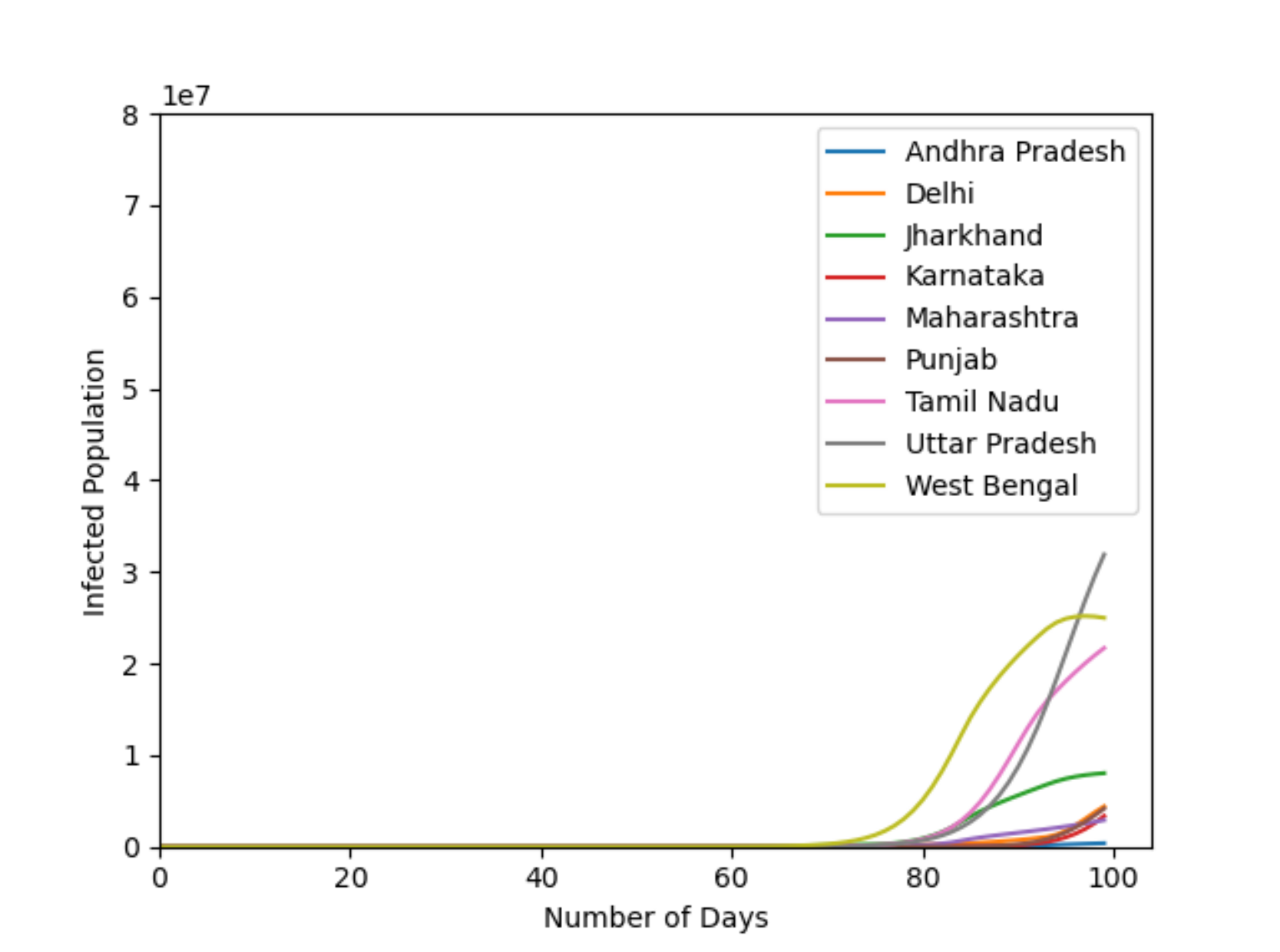}\par
    \caption{Threshold curve - II}
    \label{fig 7b}
\end{multicols}
\caption*{\textbf{  Number of infections of different states for 100 days using optimal policy}}
    \label{fig 7}
\end{figure*}

The effect of the optimal policy on the cost to economy and the number of infected people is shown in figures \ref{fig 6a}, \ref{fig 6b}, \ref{fig 7a} and \ref{fig 7b}. The optimal policy clearly has a lower economic costs when no lockdown is applied or a complete lockdown is applied. On the other hand the number of infections lie between the two extreme cases demonstrating the effect of competing objectives. 

Moreover, when we modelled social distancing with threshold curve-I, it resulted in a lower infection rate and a slightly higher cost when compared to social distancing modelled with Threshold curve II.Thus, an optimal policy can help reducing economic costs while controlling the number of infections.

\section{Limitations and future work}

\subsection{Limitations}
We consider the three major direct costs to economy during a pandemic, viz. medical cost consisting of, special life support equipment cost and medical treatment cost inclusive of testing, work absenteeism cost encountered due to work absentees of infected population, and economic sector wise cost due to either government imposed or threshold social distancing. Cost components encountered due to lifestyle changes such as reduced consumption of certain types of goods and services and other indirect costs are not considered for simplicity. We further do not consider the effect of delays in economic recovery as well and other indirect costs as well.

According to a WHO news report April, 2020 \cite{whobrief}, due to lack of evidence, it is currently unclear if recovered population that have antibodies is protected from second infection. We assume recovered population to develop immunity and not get infected again.
It should be noted that, SEIR is a generic infectious disease model and is not specific to COVID-19. Finally, there exist inherent limitations of deterministic models since these do not consider of time variance of infectivity and recovery rates \cite{roberts2015nine}.

\subsection{Future work}
Better economic models can help get us more accurate results, such as use of the Cobb-Douglas production function to estimate production value, so that the effect of the pandemic on the production value could be measured. The characteristics of COVID-19 as well as its transmission remains unknown therefore any developments could help us drive the model to include more agreed upon parameter values.

\section{Conclusion}

Given the rapid spread of COVID-19 across the globe and countries imposing severe lockdown measures to fight the virus. It has become essential to come up with a balance between cost imposed by the pandemic to the economy and the number of infections so that medical facilities are not overwhelmed. Over the basic SEIR model, we have considered  interstate movement of population and have built a strategy to obtain optimal policy solutions. We have also modelled the fraction of population that willingly follows social distancing despite no government imposed lockdowns which makes the model coherent with COVID-19 situation. The optimal policy  solution was generated by using the NSGA-II which offers a possible solution involving partial lockdown varying with time and place on the basis of infected population and cost incurred.  

\section{Acknowledgements}

\noindent We would like to thank Arun Ramamurthy for assisting us with initial discussions. We would like to acknowledge Arun’s contribution with NSGA-II optimization. Arun is a researcher at Tata Research Design and Development Center, Pune. We also want to thank Dr. Arnab Chatterjee (Scientist at TCS Innovation Labs, Delhi) and Dr. Shirish Karande (Senior Scientist at Tata Research Design and Development Center, Pune)  for their valuable evaluations and suggestions.

\bibliographystyle{plain}
\bibliography{main}

  \vspace{-5cm}
     
\end{document}